\documentclass{article}

\usepackage{arxiv}

\usepackage[utf8]{inputenc} % allow utf-8 input
\usepackage[T1]{fontenc}    % use 8-bit T1 fonts
\usepackage{hyperref}       % hyperlinks
\usepackage{bm}
\usepackage{float}
\usepackage{url}            % simple URL typesetting
\usepackage{booktabs}       % professional-quality tables
\usepackage{amsfonts}       % blackboard math symbols
\usepackage{amsmath}
\usepackage{nicefrac}       % compact symbols for 1/2, etc.
\usepackage{microtype}      % microtypography
\usepackage{lipsum}		    % Can be removed after putting your text content
\usepackage{graphicx}
\usepackage{natbib}
\usepackage{doi}
\usepackage{tikz}
\usepackage{adjustbox}
\usetikzlibrary{positioning, arrows.meta, calc, shapes}
\usepackage[section]{placeins}

\title{Detecting Narrative Shifts through Persistent Structures: A Topological Analysis of Media Discourse}

\author{\hspace{1mm}Mark M. Bailey\thanks{Corresponding author} \\
	Department of Cyber Intelligence and Data Science\\
    Director, Biological and Computational Intelligence Center\\
	National Intelligence University\\
	Bethesda, MD. USA \\
	\texttt{mark.m.bailey@ni-u.edu} \\
	\And
	\hspace{1mm}Mark I. Heiligman \\
	Faculty Emeritus\\
    Department of Cyber Intelligence and Data Science\\
	National Intelligence University\\
	Bethesda, MD. USA \\
    \texttt{mark.heiligman@gmail.com} \\
}

\date{}

\renewcommand{\shorttitle}{A Topological Analysis of Media Discourse}

\hypersetup{
pdftitle={Detecting Narrative Shifts through Persistent Structures: A Topological Analysis of Media Discourse},
pdfsubject={cs.SI, cs.CL, physics.soc-ph},
pdfauthor={Mark M. Bailey, Mark I. Heiligman},
pdfkeywords={Persistent Homology, Topological Data Analysis (TDA), Narrative Structure, Unsupervised Event Detection, Semantic Graphs, Topology of Language, Media Discourse, Computational Social Science},
}

\begin{document}
\maketitle

\begin{abstract}
How can we better detect when global events fundamentally reshape public discourse? In this study, we introduce a topological framework for tracking structural change in media narratives using persistent homology. Drawing on a corpus of international news articles spanning multiple globally significant events—including the Russian invasion of Ukraine (February 2022), the murder of George Floyd (May 2020), the U.S. Capitol insurrection (January 2021), and the Hamas-led invasion of Israel (October 2023)—we construct daily co-occurrence graphs of noun phrases to represent the evolving shape of discourse. Each graph is embedded and transformed into a persistence diagram via a Vietoris–Rips filtration. We then compute Wasserstein distances and measure persistence entropies across homological dimensions to measure semantic disruption and structural volatility over time. Our findings reveal that major geopolitical and social events correspond to sharp spikes in both $H_0$ (connected components) and $H_1$ (loops), suggesting sudden reorganization in narrative coherence and framing. Cross-correlation analyses show a characteristic lag structure in which disruptions to component-level structure typically precede changes in higher-order motifs, suggesting a bottom-up cascade of semantic reconfiguration. However, we also identify a notable exception: during the Russian invasion of Ukraine, this causal direction was reversed, with $H_1$ entropy leading $H_0$, which may suggest a top-down imposition of narrative framing before local discourse adjusted. Persistence entropy further distinguishes focused from diffuse narrative regimes. These results demonstrate that persistent homology offers a mathematically grounded, unsupervised method for detecting inflection points and directional shifts in public attention, without requiring prior event annotation. This approach provides new tools for computational social science, enabling real-time detection and interpretation of semantic restructuring during crises, protests, and information shocks in global discourse.
\end{abstract}


\section{Introduction}
In the wake of globally significant events—wars, invasions, pandemics, social movements—the structure of public discourse undergoes rapid and often radical transformation. News media, as both sensor and amplifier of collective attention, does not merely report events but actively reorganizes the social understanding of what matters, who matters, and how narratives interrelate. These shifts are not always visible in individual headlines, but can manifest as deeper structural changes in the way language coalesces around people, ideas, and events.

Computational social scientists have long sought to detect such inflection points using a variety of techniques: keyword frequency \citep{church1995poisson,pennebaker2015liwc}, sentiment analysis \citep{pang2008opinion,liu2012sentiment}, topic modeling \citep{blei2003latent,roberts2014structural}, and vector-based representations of meaning \citep{mikolov2013efficient,pennington2014glove}. While these methods have yielded important insights, they tend to focus on lexical content in isolation—emphasizing what is being discussed over how discourse is structured. In particular, they often struggle to capture higher-order dependencies, thematic loops, or shifts in the underlying geometry of discourse networks.

Recent advances in topological data analysis (TDA) offer a compelling alternative. Persistent homology, in particular, has emerged as a powerful method for characterizing multiscale structure in complex systems \citep{carlsson2009topology,wasserman2018tda}. In network neuroscience, persistent homology has revealed non-trivial loop structures critical for understanding functional connectivity \citep{sizemore2019importance}, while in dynamic networks, it has been used to detect temporal structural shifts \citep{stolz2017persistent,myers2023}. Foundational work has also established the mathematical stability of persistence diagrams and entropy metrics \citep{cohen2007stability,chazal2016structure,atienza2020persistent,atienza2016separating}.

In natural language processing, persistent homology has been used to represent the structural features of text \citep{Zhu2013}, to differentiate literary styles \citep{Gholizadeh2018}, and to capture attention patterns in transformer models \citep{Kushnareva2021}. Other work has applied TDA to document summarization \citep{Haghighatkhah2022} and discourse semantics \citep{Savle2019}, while broader NLP applications of TDA include narrative shift detection and media structure analysis \citep{Rocha2024,Arun2025,Nguyen2020}.

In this study, we propose a new approach: to detect global narrative change by measuring the topological structure of media discourse as it evolves over time. Drawing on tools from algebraic topology—specifically, persistent homology—we analyze graphs constructed from co-occurring noun phrases in news coverage. Each graph represents a snapshot of discourse for a particular day, with edges reflecting contextual proximity between key terms. By embedding these graphs in geometric space using Word2Vec \citep{mikolov2013efficient}, reducing dimensionality using UMAP \citep{mcinnes2018umap}, and applying a Vietoris–Rips filtration \citep{zomorodian2005computing,maria2014gudhi}, we extract persistence diagrams that encode the birth and death of connected components and cycles—topological proxies for the cohesion and framing of public narratives. Our method offers a general framework for event detection that does not rely on predefined keywords, topics, or training data, but instead captures the geometry of collective meaning-making as it unfolds.

Our motivation is rooted in the belief that events of global salience—those that reorder geopolitical realities—will also reorder the structure of discourse itself. To test this, we examine news coverage surrounding four major crises: the Russian invasion of Ukraine on February 24\textsuperscript{th}, 2022, the Hamas-led invasion of Israel on October 7\textsuperscript{th}, 2023, the murder of George Floyd on May 25\textsuperscript{th}, 2020, and the U.S. Capitol insurrection on January 6\textsuperscript{th}, 2021. We retrieve articles using NewsAPI \citep{newsapi} and analyze the resulting daily co-occurrence networks to track changes in topological structure.

We build on prior research that has modeled narrative dynamics through temporal semantic graphs \citep{Opdahl2023,Yan2023,Radicioni2021} and unsupervised event detection techniques \citep{Zhao2023,Hajij2018,Aktas2019}. Our use of persistent homology aligns with recent efforts to detect political controversy and network anomalies through topological signatures \citep{Rocha2024,Arun2025,Nguyen2020}.

This work contributes to computational social science by introducing a new class of topological metrics for understanding narrative evolution. It complements existing linguistic and network-based methods with a structural, scale-invariant tool set for detecting rupture, drift, and reorganization in the public sphere. As computational models of society increasingly incorporate complexity and emergence \citep{may2008ecology,guilbeault2021topological}, persistent homology offers a powerful lens for studying how large-scale meaning structures form, dissolve, and reassemble in response to the shocks of history.

\section{Theoretical Foundations}
\subsection{Filtrations and Persistence}
Let \(X \subseteq \mathbb{R}^d\) be a finite point cloud. To analyze the topology of \(X\), we construct a Vietoris–Rips filtration \(\{R_\epsilon(X)\}_{\epsilon \geq 0}\), where each complex \(R_{\epsilon}(X)\) includes a \(k\)-simplex whenever all pairwise distances between its \(k+1\) vertices are within distance \(\epsilon\):

\begin{equation}
\mathcal{R}_\epsilon(X) = \left\{ \sigma \subseteq X \;\middle|\; \|x_i - x_j\| \leq \epsilon,\; \forall x_i, x_j \in \sigma \right\}.
\end{equation}

As $\epsilon$ increases, new topological features (connected components, loops, voids) are born and others disappear, forming a topological fingerprint of the data's multiscale structure. These features are recorded in persistence diagrams as points \((b_i,d_i) \in \mathbb{R}^2\), where \(b_i\) and \(d_i\) represent the birth and death scale of a feature in homology group \(H_k\):

\begin{equation}
D_k(X) = \left\{ (b_i, d_i) \ \middle| \ \text{feature } i \text{ in homology group } H_k \right\}, \quad \text{ with } d_i > b_i.
\end{equation}

We illustrate the filtration process and the resulting diagrammatic representation in Figure \ref{fig:rips_filtration}:

\begin{figure}[!htbp]
\centering
\begin{tikzpicture}[scale=1.2, every node/.style={circle, fill=black, inner sep=1pt}]
  % Initial point cloud
  \node[label=left:{$x_1$}] (x1) at (0,0) {};
  \node[label=above:{$x_2$}] (x2) at (2,2) {};
  \node[label=right:{$x_3$}] (x3) at (4,0) {};
  \node[label=below:{$x_4$}] (x4) at (2,-2) {};

  % Draw epsilon balls
  \foreach \x in {x1,x2,x3,x4} {
    \node[draw=gray, fill=gray!20, minimum size=2.25cm, circle, opacity=0.3] at (\x) {};
  }

  % Draw epsilon arrow
  \draw[dotted, ->] (x2) -- (3,2) node[inner sep=0pt, minimum size=0pt, midway, above, label={\( \epsilon \)}]{};

  % Draw edges for epsilon small
  \draw[thick] (x1) -- (x2);
  \draw[thick] (x1) -- (x4);
  \draw[thin] (x2) -- (x3);
  \draw[thin] (x3) -- (x4);

  % Simplices at larger epsilon
  \filldraw[fill=black!10, thick, draw=black] (x1.center) -- (x2.center) -- (x4.center) -- cycle;
  \draw[thick, black] (x1) -- (x2);
  \draw[thick, black] (x2) -- (x4);
  \draw[thick, black] (x1) -- (x4);

  \node[draw=none, fill=none] at (2, -3.5) {\small Example of Vietoris–Rips filtration};

\end{tikzpicture}
\vspace{-5em}
\caption{Schematic of a Vietoris–Rips filtration on a point cloud. As the scale parameter \( \epsilon \) increases, points become connected into edges, triangles, and higher-dimensional simplices. Persistent homology tracks the birth and death of topological features across this growing sequence of simplicial complexes.}
\label{fig:rips_filtration}
\end{figure}

\subsection{Normalized Wasserstein Distance}
To quantify how topological structures evolve over time, we compute the Wasserstein distance (a.k.a. "earth mover's distance") between persistence diagrams of successive graph windows. The \(p\)-Wasserstein distance between two persistence diagrams \( D_1 \) and \( D_2 \) is defined as:

\begin{equation}
W_p(D_1, D_2) = \left( \inf_{\gamma : D_1 \to D_2} \sum_{x \in D_1} \| x - \gamma(x) \|_q^p \right)^{1/p},
\end{equation}

where \( \gamma \) ranges over all bijections between the two diagrams, \( \| \cdot \|_q \) is the \( \ell^q \)-norm, and \( p \) is typically 1 or 2.

To ensure comparability across time, we used the normalized Wasserstein distance:

\begin{equation}
\widetilde{W}_p(D_1, D_2) = \frac{W_p(D_1, D_2)}{|D_1| + |D_2|},
\end{equation}

This accounts for differences in the number of points in each diagram.

\subsection{Normalized Persistence Entropy}
We also computed persistence entropy to examine the distributional complexity. For a given persistence diagram \( D \), let \( \ell_i = d_i - b_i \) be the lifetime of the \( i \)-th feature born at \( b_i \) and dying at \( d_i \). The normalized persistence entropy is defined as:

\begin{equation}
\widetilde{H}(D) = - \frac{1}{\log n}\sum_{i=1}^n \frac{\ell_i}{\sum_j \ell_j} \log \left( \frac{\ell_i}{\sum_j \ell_j} \right),
\end{equation}

where the sum runs over all points with \( \ell_i > 0 \). This provides a measure of the distributional complexity of lifetimes in the topological signature.

\section{Methodology}
\subsection{Data Collection and Pre-processing}
To analyze structural changes in global news discourse over time, we began by collecting a corpus of daily news articles using the NewsAPI for a date range bracketing the event of interest.\citep{newsapi} Articles were drawn from several national and international media outlets spanning the political spectrum (see Table \ref{tab:news_sources}) and text was aggregated by publication date. Using standard natural language processing (NLP) techniques, we lemmatized the text and extracted noun phrases from each article.\citep{loper-bird-2002-nltk} For every day \( t \), we constructed an undirected graph \( G_t = (V_t, E_t) \), where nodes \( V_t \) represent unique nouns or adjectives and edges \( E_t \) encode co-occurrence within the same extracted phrase.

\begin{table}[H]
\centering
\caption{News Sources Queried via NewsAPI}
\label{tab:news_sources}
\begin{tabular}{lll}
\toprule
\multicolumn{3}{c}{\textbf{Sources}} \\
\midrule
ABC News & Al Jazeera English & Associated Press \\
Axios & Bloomberg & Business Insider \\
CBS News & CNN & ESPN \\
Fox News & Google News & Hacker News \\
IGN & MSNBC & National Geographic \\
National Review & NBC News & New Scientist \\
Newsweek & New York Magazine & Next Big Future \\
Politico & Recode & Reddit /r/all \\
Reuters & TechCrunch & TechRadar \\
The American Conservative & The Hill & The Huffington Post \\
The Next Web & The Wall Street Journal & The Washington Post \\
The Washington Times & Time Magazine & USA Today \\
Vice News & Wired & \\
\bottomrule
\end{tabular}
\end{table}

\subsection{Graph Construction and Embedding Generation}
To embed each graph in a semantic vector space while preserving its local neighborhoods and meso-scale community structure, we created a time series of daily co-occurrence graphs \( \{G_1, G_2, \dots, G_T\} \), which were aggregated into overlapping windows of size \( w \) and stride \( s \) to induce temporally smoothed subgraphs. We define the \( k \)-th windowed graph \( \mathcal{G}_k \) as the union of \( w \) consecutive graphs starting at index \( (k - 1)s + 1 \):

\begin{equation}
\mathcal{G}_k = \bigcup_{t = (k - 1)s + 1}^{(k - 1)s + w} G_t
\end{equation}

\begin{figure}[!htbp]
\centering
\begin{tikzpicture}[node distance=1.8cm, every node/.style={font=\small}]
  % Graph section
  \node[draw, circle] (A) at (0,0) {A};
  \node[draw, circle] (B) at (1.5,1.2) {B};
  \node[draw, circle] (C) at (3,0) {C};
  \node[draw, circle] (D) at (4.5,1.2) {D};
  \node[draw, circle] (E) at (6,0) {E};

  \draw[->, thick] (A) -- (B);
  \draw[->, thick] (B) -- (C);
  \draw[->, thick] (C) -- (D);
  \draw[->, thick] (D) -- (E);

  \node at (3,2) {\textbf{Random Walk Path:} A $\rightarrow$ B $\rightarrow$ C $\rightarrow$ D $\rightarrow$ E};

  % Word2Vec section
  \node[rectangle, draw, minimum width=5cm, minimum height=1.2cm, anchor=north west] (w2v) at (3.0,-1.3) {
    \begin{minipage}{5.9cm}
      \centering
      Word2Vec training window \\
      (e.g., center = C, context = B and D)
    \end{minipage}
  };

  % Arrows from path to W2V
  \draw[->, dotted] (A) -- (8,1.3);
  \draw[->, dotted] (B) -- (9.1,1.7);
  \draw[->, dotted] (C) -- (9.1,1.0);
  \draw[->, dotted] (D) -- (9.9,1.1);
  \draw[->, dotted] (E) -- (10.5,1.6);

  % Embedding space
  \node at (9.75,-0.3) {\textbf{Embedding Space (2D projection)}};
  \draw[->] (7.5,0) -- (12.5,0);
  \draw[->] (7.5,0) -- (7.5,2.5);

  \node[circle, draw, fill=black, inner sep=1pt, label=right:{\small A}] at (8,1.3) {};
  \node[circle, draw, fill=black, inner sep=1pt, label=right:{\small B}] at (9.1,1.7) {};
  \node[circle, draw, fill=black, inner sep=1pt, label=right:{\small C}] at (9.1,1.0) {};
  \node[circle, draw, fill=black, inner sep=1pt, label=right:{\small D}] at (9.9,1.1) {};
  \node[circle, draw, fill=black, inner sep=1pt, label=right:{\small E}] at (10.5,1.6) {};

  \draw[->, thick] (6,-1.3) -- (7.5,0);

\end{tikzpicture}
\caption{Word2Vec maps graph paths to continuous embeddings by learning node co-occurrence within sampled walks. Paths like A–B–C–D–E are treated as sentences, and Word2Vec learns that nodes like B, C, and D tend to appear in similar contexts. The result is a spatial representation in which proximity reflects semantic and structural similarity.}
\label{fig:word2vec_diagram}
\end{figure}
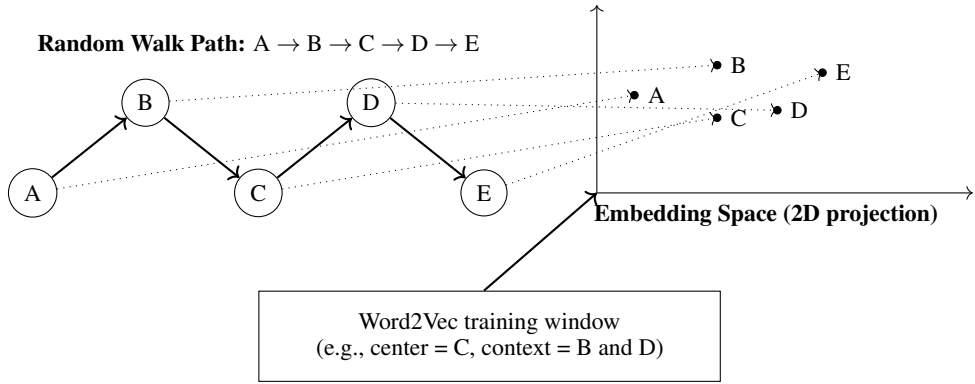

This formulation yields a temporally smoothed representation of discourse structure, where each \( \mathcal{G}_k \) subgraph encodes the latent topological features present across a contiguous window of days. The stride \( s \) controls the overlap between successive windows: when \( s < w \), the windows overlap; when \( s = w \), the windows are disjoint. To reduce noise, we removed all nodes of degree one. The resulting graphs capture the structural backbone of daily discourse—highlighting the most contextually significant and interconnected concepts. For a visual representation of the embedding process, see Figure \ref{fig:word2vec_diagram}

A flattened graph across the entire time series, \( G_{\text{flat}} = \bigcup_{t=1}^T G_t \), was also constructed and used to train a global Word2Vec embedding model.\citep{mikolov2013word2vec} Random walks through \( G_{\text{flat}} \) were treated as pseudo-sentences to train the model, and embeddings were then used to compute edge vectors for each windowed graph. This method mirrors the logic of the Node2Vec algorithm but provides greater control and interpretability by decoupling walk generation from embedding.\citep{grover2016node2vec} 

The Word2Vec model trained using the following hyperparameters:

\begin{itemize}
    \item Number of walks (training): 100,000
    \item Number of walks (inference): 4,000
	\item Dimensions: 64
	\item Window size: 3
    \item Stride: 1
    \item Walk length: 40
	\item Epochs: 10
    \item Skip-gram model (sg = 1)
\end{itemize}

Edge embeddings were then derived from node embeddings by averaging the Word2Vec vectors of each edge’s endpoints:

\begin{equation}
\mathbf{e}_{ij} = \frac{1}{2} \left( \mathbf{v}_i + \mathbf{v}_j \right)
\end{equation}

where $\mathbf{v}_i$ and $\mathbf{v}j$ are the Word2Vec embeddings for nodes $i$ and $j$, respectively. The resulting point cloud ${\mathbf{e}{ij}}$ forms a geometric representation of the latent structure of daily discourse.

Unlike frequency-based co-occurrence matrices, Word2Vec embeddings learn distributed representations that capture nuanced relationships across local and global graph neighborhoods. By training Word2Vec on random walks through noun phrase co-occurrence graphs, we embed nodes in a continuous space where semantic and structural similarity are encoded in geometric proximity. This embedding serves as a soft projection of the discourse topology, with edge vectors encoding both lexical co-dependence and relational structure. Crucially, this continuous representation facilitates persistent homology by transforming the graph into a metric space suitable for Rips filtration.

\subsection{Dimensionality Reduction}
We reduced the high-dimensional edge embeddings to two dimensions using Uniform Manifold Approximation and Projection (UMAP), chosen for its ability to preserve local topological structure.\citep{mcinnes2018umap} This 2D embedding of edges serves as a geometric point cloud capturing the structure of daily discourse.

\subsection{Persistent Homology and Topological Feature Extraction}
Persistent homology is a foundational method in topological data analysis (TDA) used to extract and quantify the shape of data across multiple scales. Rather than fixing a single neighborhood or connectivity threshold, persistent homology operates over a filtration: a growing sequence of simplicial complexes that gradually “connect” the data. This multiscale approach reveals features such as connected components, loops, and voids that persist across a range of proximity values.

In this analysis, we applied a Rips filtration using the Python Gudhi library, constructing simplicial complexes over the embedded edge points and computing persistence diagrams for homology dimensions 0 and 1.\citep{gudhi:2021, maria2014gudhi} This allowed us to extract topological features, specifically connected components (in \(H_0\)) and loops (in \(H_1)\).

Persistent homology enjoys a critical stability theorem: small perturbations in the input data (such as minor differences in edge embeddings due to random walk sampling) produce only small changes in the persistence diagrams, under both bottleneck and Wasserstein distances. \citep{carlsson2009topology, chazal2016structure} This makes the method robust to noise and highly suitable for extracting meaningful patterns from high-dimensional, real-world data such as language. In this study, we interpret long-lived topological features as persistent semantic or narrative structures, while short-lived features are treated as noise. Shifts in these diagrams correspond to inflection points in discourse topology - moments when the structure of news coverage reorganizes in response to real-world events.

\subsection{Wasserstein Distance and Persistence Entropy Analysis}
We then computed Wasserstein distances between the persistence diagrams of successive windows to measure the magnitude of topological change between time steps. This yielded a time series of distances for both \(H_0\) and \(H_1\), where spikes correspond to abrupt reorganizations in the structure of daily discourse. The distance traces were smoothed and first and second derivatives calculated. We also computed persistence entropy for each window, providing a scalar summary of the distribution of topological feature lifetimes. High entropy indicates many short-lived features (suggesting noise or narrative diffusion), while low entropy reflects dominant, persistent structures (suggesting focused narrative framing). Entropy traces were smoothed and first and second derivatives calculated.

\subsection{Interpreting Topological Signals}
\paragraph{Persistence Entropy.} 
Persistence entropy quantifies the informational complexity of a persistence diagram by summarizing the distribution of feature lifetimes (i.e., the differences between birth and death times of topological features). A high entropy value indicates a relatively uniform distribution of lifetimes, suggesting a fragmented or diffuse topological structure in the underlying semantic graph. This may correspond to a multipolar discourse in which many competing themes coexist. Conversely, low entropy reflects the dominance of a few long-lived features, implying a more coherent or thematically unified narrative structure.

\paragraph{First Derivative of Persistence Entropy.}
The first temporal derivative of persistence entropy captures the rate at which structural complexity in the discourse is changing. Positive spikes in this derivative indicate moments of narrative disruption or diversification—typically associated with the onset of new, competing topics or conceptual instability. Negative spikes, in contrast, suggest narrative convergence, simplification, or the formation of dominant thematic structures.

\paragraph{Second Derivative of Persistence Entropy.}
The second derivative measures the acceleration or deceleration in the rate of topological change. Sustained positive values indicate growing volatility in discourse structure—potentially corresponding to escalating uncertainty or the rapid proliferation of competing framings. Sustained negative values suggest a deceleration of change, marking a transition toward stabilization or the emergence of a new semantic regime. Inflection points in this signal are often interpretable as phase transitions in public discourse.

\paragraph{Wasserstein Distance Between Persistence Diagrams.}
The Wasserstein distance measures the dissimilarity between two successive persistence diagrams, providing a direct quantification of topological drift over time. A large Wasserstein distance between days \( t \) and \( t+1 \) indicates a significant reorganization in the topology of the discourse graph—implying a structural shift in how concepts relate or cluster. This measure is sensitive to the emergence, disappearance, and transformation of semantic motifs (e.g., communities, loops). Periods with elevated Wasserstein distances often coincide with major inflection points in the narrative system, such as the immediate aftermath of a global event or a critical change in media framing.

\paragraph{Combined Interpretation.}
Taken together, these topological signals form a multiscale diagnostic framework. Persistence entropy describes the current complexity of narrative structure, its derivatives signal transitions or volatility, and Wasserstein distance quantifies structural change between states. Synchronizing these signals with known events or anomalous dates provides insight into how global discourse evolves—revealing both immediate ruptures and delayed cascades in narrative organization.

\section{Results}
This topological framework combining graph-based NLP, Word2Vec embeddings, persistent homology, and time-series analysis offers a scalable and interpretable method for detecting structural transitions in complex semantic data streams such as news discourse.

\subsection{October 7th, 2023}
\begin{figure}[htbp]
    \centering
    \includegraphics[width=\linewidth]{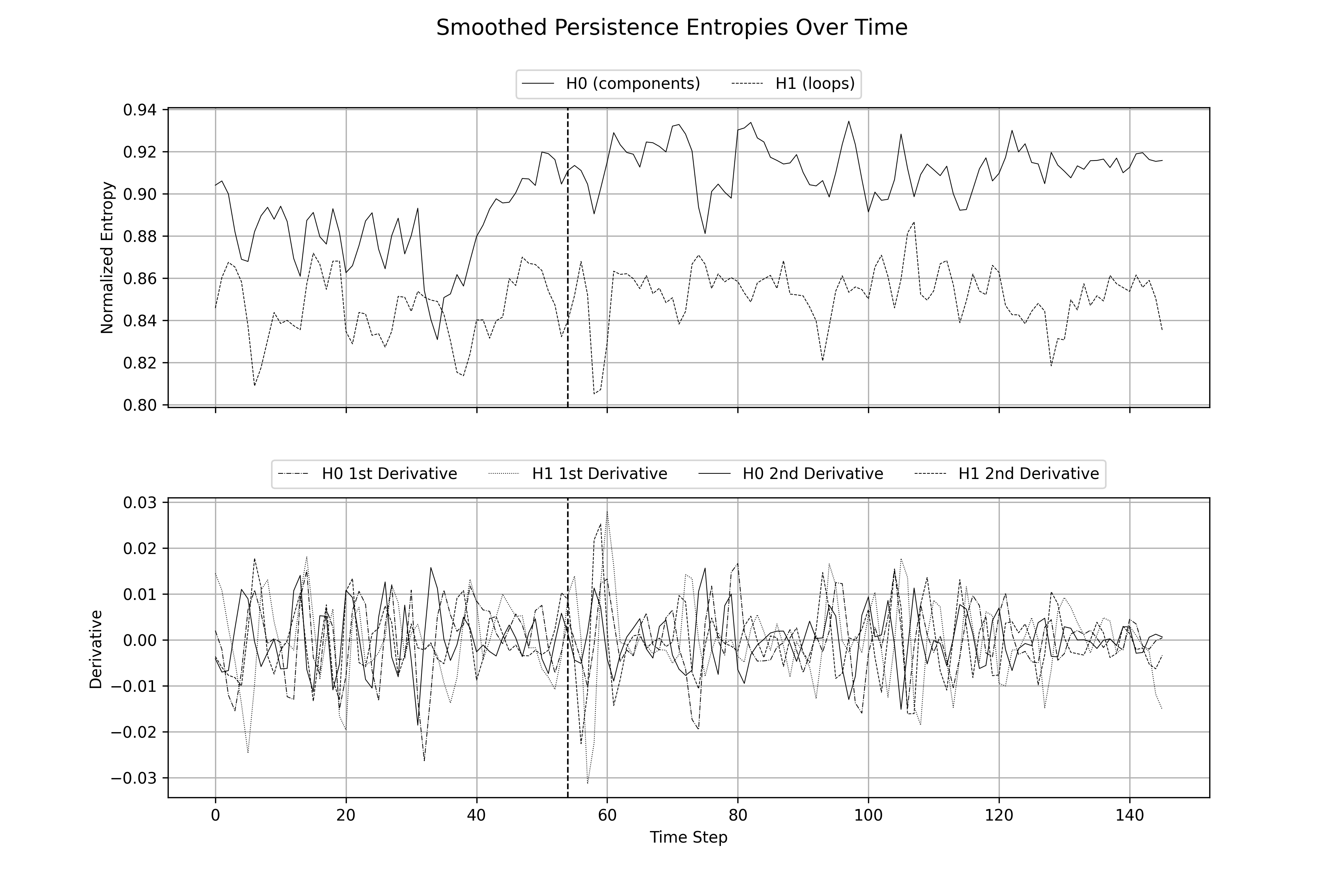}
    \caption{\textbf{October 7\textsuperscript{th} Dataset: Persistence Entropies.} Smoothed persistence entropies (top panel) and their first and second derivatives (bottom panel) for homology dimensions 0 ($H_0$) and 1 ($H_1$). A sharp rise in $H_0$ entropy occurs immediately preceding the marked event (dashed line – October 7\textsuperscript{th}, 2023), while $H_1$ entropy remains relatively stable or declines. The derivative traces reveal pronounced first- and second-order changes in $H_0$, identifying the transition point as a dynamic regime shift in the graph’s connected component structure. This suggests an abrupt increase in the disorder or complexity of the system’s zero-dimensional topology, followed by stabilization. The contrast in behavior between $H_0$ and $H_1$ suggests different roles for component and loop features in driving structural change.}
    \label{fig:o_7_entropy_derivatives}
\end{figure}

\begin{figure}[htbp]
    \centering
    \includegraphics[width=\linewidth]{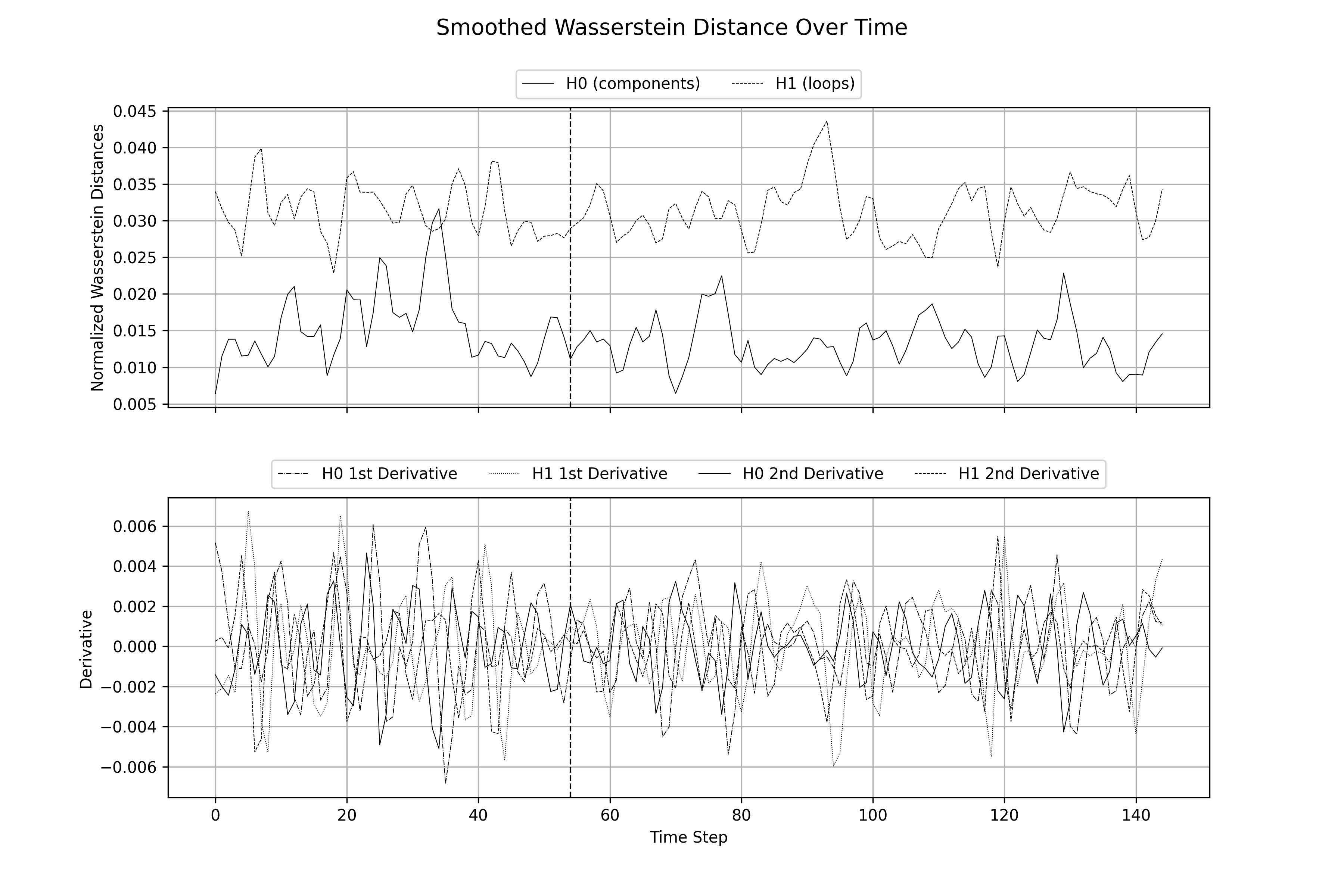}
    \caption{\textbf{October 7\textsuperscript{th} Dataset: Wasserstein Distance.} Smoothed, normalized Wasserstein distances over time (top panel) and their first and second temporal derivatives (bottom panel), for persistence diagrams in homology dimensions 0 ($H_0$, connected components) and 1 ($H_1$, loops). The vertical dashed line denotes a known event (October 7\textsuperscript{th}, 2023). While raw distance values change gradually, the derivative panel suggests a marked transition: both first- and second-order derivatives converge around zero post-event, suggesting a phase of topological stabilization following the disruption. This indicates that dynamic variability in persistence diagrams decreases significantly after the event, even if the absolute topological distances remain subtle.}
    \label{fig:o_7_distance_derivative}
\end{figure}

\begin{figure}[htbp]
    \centering
    \includegraphics[width=\linewidth]{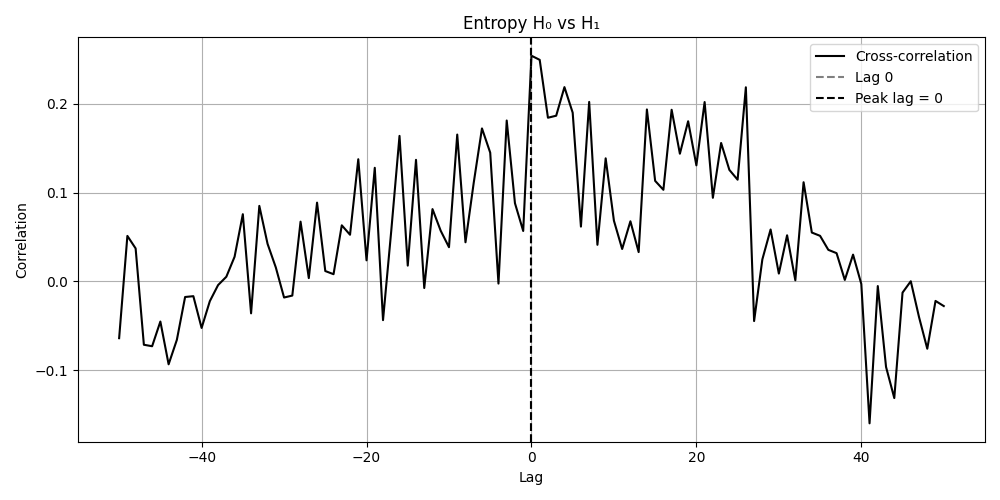}
    \caption{\textbf{October 7\textsuperscript{th} Dataset: Persistence Entropy Cross Correlation.} Cross-correlation between $H_0$ and $H_1$ persistence entropies. The peak correlation occurs at lag $0$, indicating that changes in entropy for connected components ($H_0$) and loops ($H_1$) are temporally synchronized. This suggests that topological complexity across homology dimensions tends to evolve concurrently, reflecting simultaneous shifts in both macro- and micro-level structure in the underlying noun-phrase co-occurrence graphs.}
    \label{fig:o_7_entropy_xcorr}
\end{figure}

\begin{figure}[htbp]
    \centering
    \includegraphics[width=\linewidth]{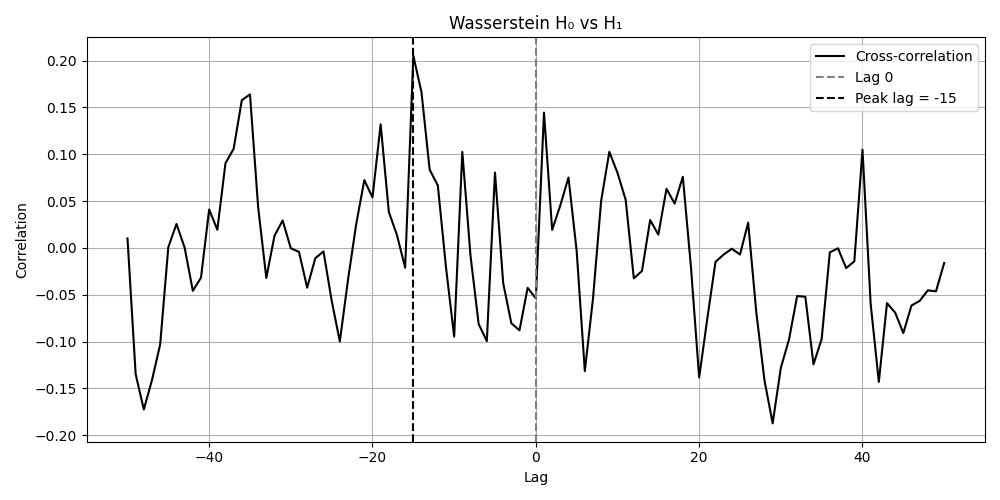}
    \caption{\textbf{October 7\textsuperscript{th} Dataset: Wasserstein Distance Cross Correlation.} The peak correlation is observed at lag $-15$, indicating that changes in $H_0$ (connected components) precede changes in $H_1$ (loops) by approximately 15 time steps. This temporal offset implies a cascading dynamic, where large-scale structural reorganization occurs before local cyclic features are affected, potentially reflecting a sequential response to systemic perturbations or shifts in narrative focus within the corpus.
    }
    \label{fig:o_7_wasserstein_xcorr}
\end{figure}

\textbf{Topological Signal Analysis of the October 7\textsuperscript{th} Dataset.} The topological signatures derived from the October 7\textsuperscript{th} dataset reveal subtle but interpretable changes in the evolving semantic structure of global news discourse. In the persistence entropy traces, a clear divergence emerges between the $H_0$ (connected components) and $H_1$ (loops) signals around the annotated event time (Figure \ref{fig:o_7_entropy_derivatives}). This divergence is mirrored in the first and second derivatives, where a spike in magnitude suggests a moment of rapid topological reorganization. In the corresponding Wasserstein distance traces, although the smoothed distance values show only minor fluctuations, the derivative plots reveal a suppression of volatility at the event marker, potentially indicating a transient stabilization in both component and cycle structure (Figure \ref{fig:o_7_distance_derivative}). Cross-correlation analysis further supports these interpretations: the entropy signals peak at lag 0, suggesting synchronous reconfiguration of component- and cycle-level entropy (Figure \ref{fig:o_7_entropy_xcorr}). In contrast, the Wasserstein signals exhibit a peak correlation at a negative lag, implying that shifts in $H_0$ structure anticipate those in $H_1$ (Figure \ref{fig:o_7_wasserstein_xcorr}). Taken together, these results indicate that the semantic network encoded in news narratives underwent a coordinated topological shift at the time of the event, with subtle precursors in component connectivity preceding more global reorganization in loop structure.

\subsection{January 6th, 2023}
\begin{figure}[htbp]
    \centering
    \includegraphics[width=\linewidth]{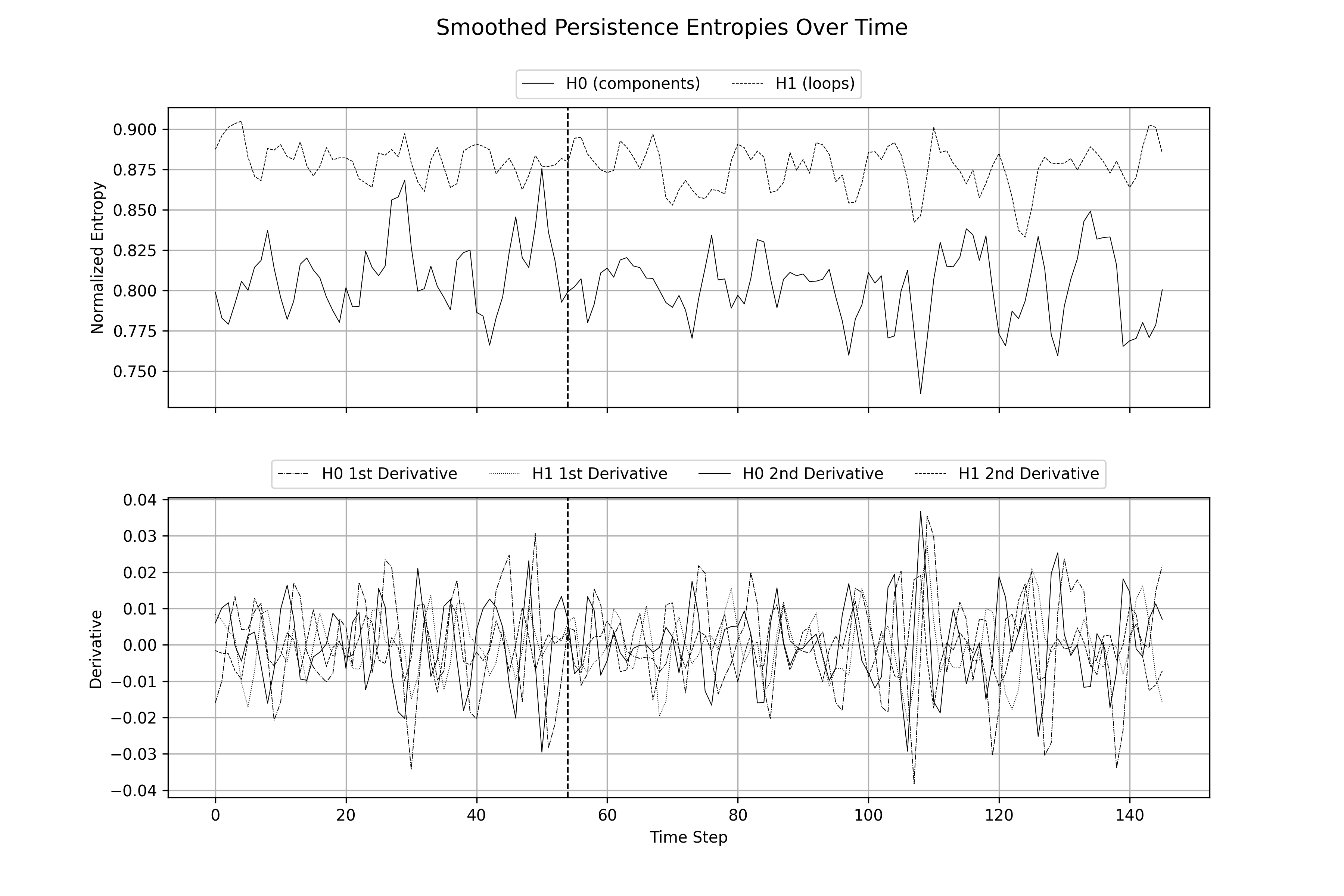}
    \caption{\textbf{January 6\textsuperscript{th} Dataset: Persistence Entropy.} Smoothed persistence entropy traces over time for $H_0$ (connected components) and $H_1$ (loops), computed from daily semantic graphs derived from global news sources. The top panel shows normalized entropy values, while the bottom panel displays first and second derivatives. A sharp rise in $H_0$ entropy and increased derivative magnitude is observed near the event time, suggesting a topological phase transition in component structure.}
    \label{fig:j_6_entropy_derivatives}
\end{figure}

\begin{figure}[htbp]
    \centering
    \includegraphics[width=\linewidth]{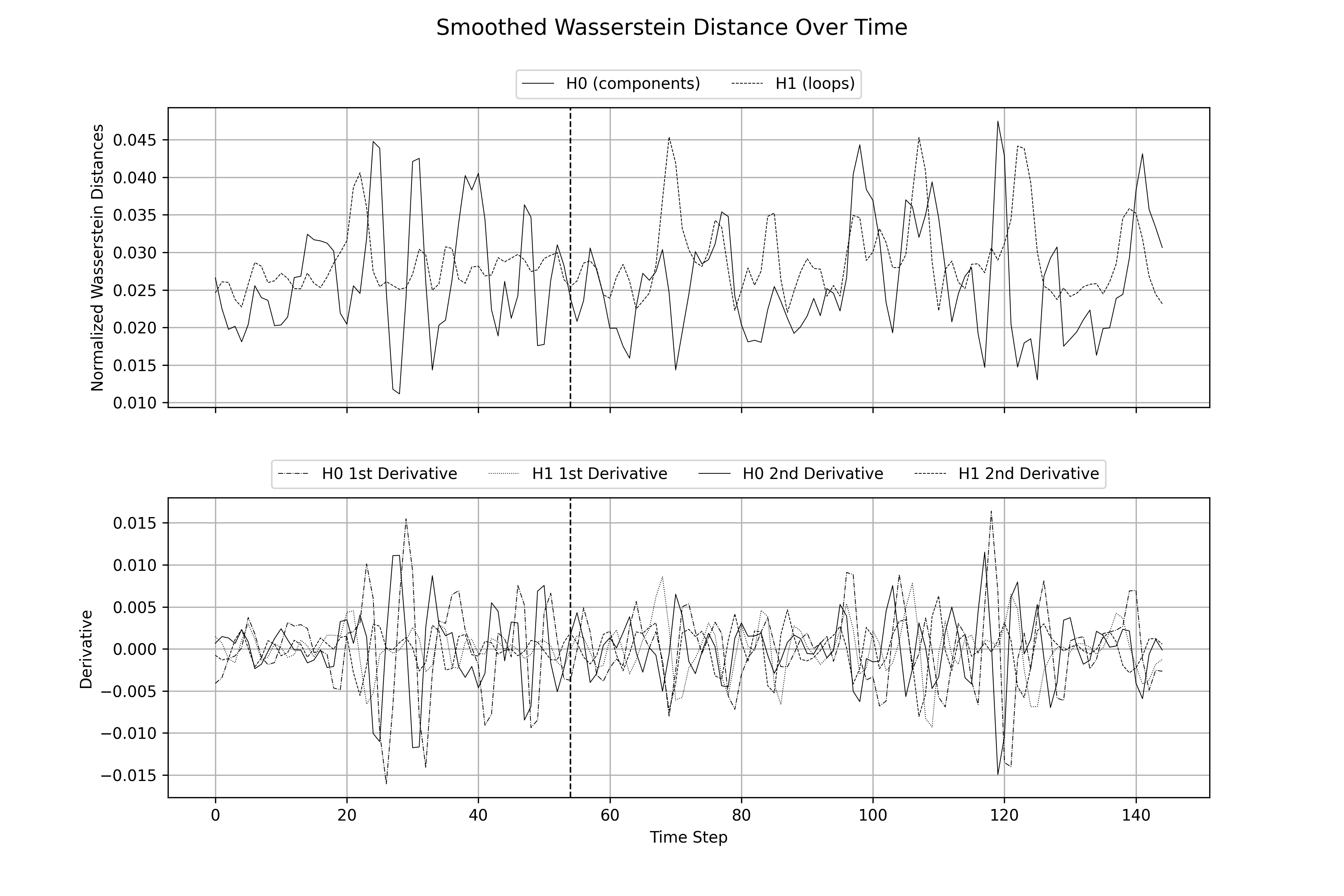}
    \caption{\textbf{January 6\textsuperscript{th} Dataset: Wasserstein Distance.} Smoothed Wasserstein distance traces over time for $H_0$ and $H_1$ persistence diagrams. The top panel shows normalized inter-diagram distances, and the bottom panel presents first and second derivatives. Prior to the event, a noticeable drop in derivative volatility suggests structural stabilization across both homology dimensions.}
    \label{fig:j_6_distance_derivatives}
\end{figure}

\begin{figure}[htbp]
    \centering
    \includegraphics[width=\linewidth]{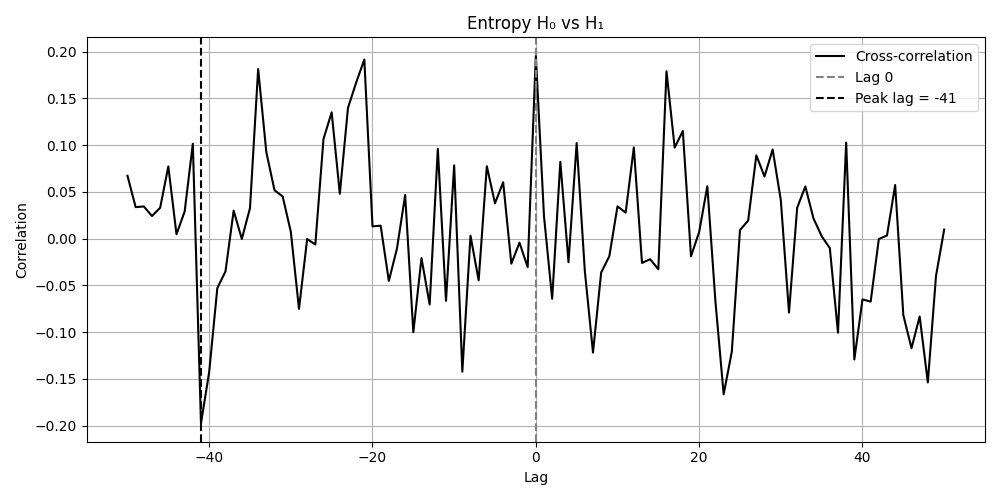}
    \caption{\textbf{January 6\textsuperscript{th} Dataset: Persistence Entropy Cross Correlation.} Cross-correlation between $H_0$ and $H_1$ persistence entropy traces. The peak correlation occurs at a lag of $-41$, indicating that changes in component-level entropy ($H_0$) precede changes in cycle-level entropy ($H_1$). This lag suggests temporal causal ordering in topological reconfiguration.}
    \label{fig:j_6_entropy_xcorr}
\end{figure}

\begin{figure}[htbp]
    \centering
    \includegraphics[width=\linewidth]{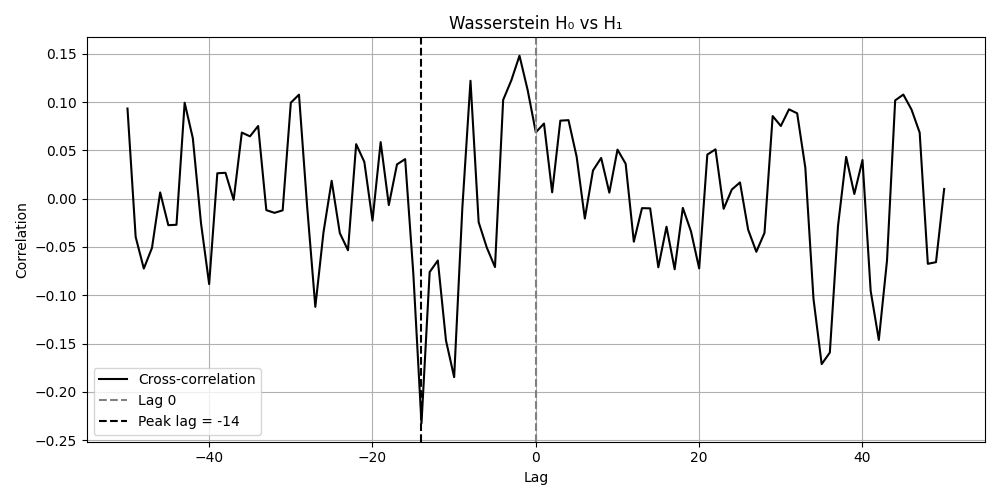}
    \caption{\textbf{January 6\textsuperscript{th} Dataset: Wasserstein Distance Cross Correlation.} Cross-correlation between $H_0$ and $H_1$ Wasserstein distance traces. The strongest correlation appears at a lag of $-14$, implying that shifts in the structure of connected components anticipate changes in loop structure. This behavior is consistent with a bottom-up reorganization of semantic topology.}
    \label{fig:j_6_wasserstein_xcorr}
\end{figure}

\textbf{Topological Signal Analysis of the January 6\textsuperscript{th} Dataset.} Analysis of topological traces around the January 6\textsuperscript{th} event revealed a clear divergence between $H_0$ and $H_1$ entropy, with $H_0$ exhibiting a sharp increase and $H_1$ remaining relatively stable (Figure \ref{fig:j_6_entropy_derivatives}). Derivative analysis further revealed transient spikes in both the first and second derivatives of $H_0$ entropy, suggesting a phase of topological acceleration preceding the event. Corresponding patterns in Wasserstein distances showed decreased volatility in the first derivative immediately before the event, indicating a period of structural consolidation (Figure \ref{fig:j_6_distance_derivatives}). Cross-correlation analysis confirmed that changes in $H_0$ consistently led those in $H_1$, with peak lag values of $-41$ for entropy and $-14$ for Wasserstein distance, suggesting a causal hierarchy wherein shifts in component connectivity precede the reorganization of higher-order cycles (Figures \ref{fig:j_6_entropy_xcorr} and \ref{fig:j_6_wasserstein_xcorr}). Together, these findings support the use of persistent homology as a sensitive tool for detecting early signs of semantic structural reconfiguration in text-derived networks.

\subsection{February 24th, 2022}
\begin{figure}[htbp]
    \centering
    \includegraphics[width=\linewidth]{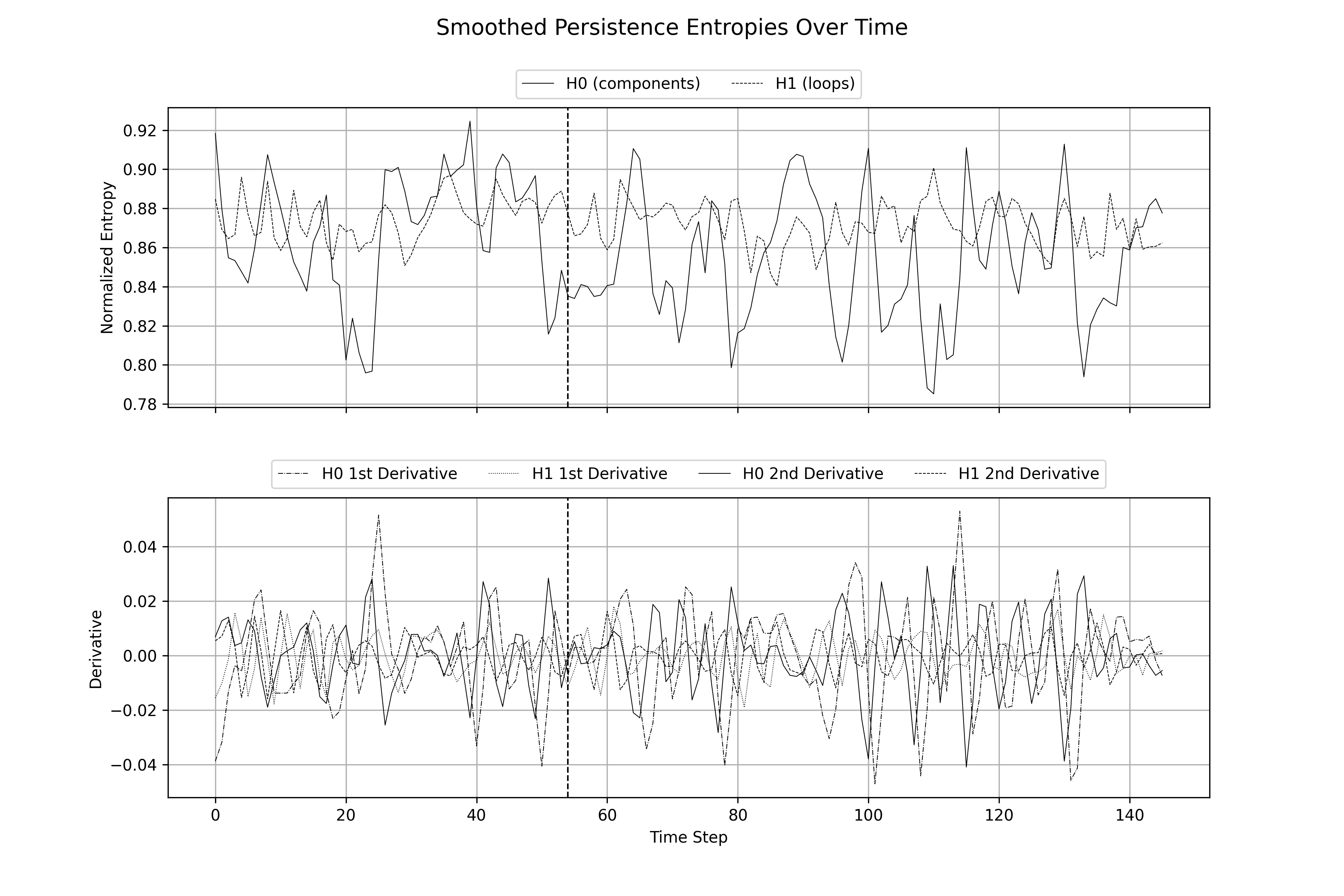}
    \caption{\textbf{February 24\textsuperscript{th} Dataset: Persistence Entropy.} Normalized persistence entropy traces (top panel) and their first and second derivatives (bottom panel) for homology dimensions $H_0$ and $H_1$. The $H_0$ entropy trace shows a sharp decline preceding the event date, followed by increased volatility and derivative acceleration—signaling a loss of structural cohesion and a shift into a more fragmented narrative state. $H_1$ entropy remains relatively steady, indicating that loop complexity was less affected. Together, these dynamics point to a breakdown and subsequent reorganization of foundational narrative elements in the lead-up to the invasion.}
    \label{fig:f_24_entropy_derivatives}
\end{figure}

\begin{figure}[htbp]
    \centering
    \includegraphics[width=\linewidth]{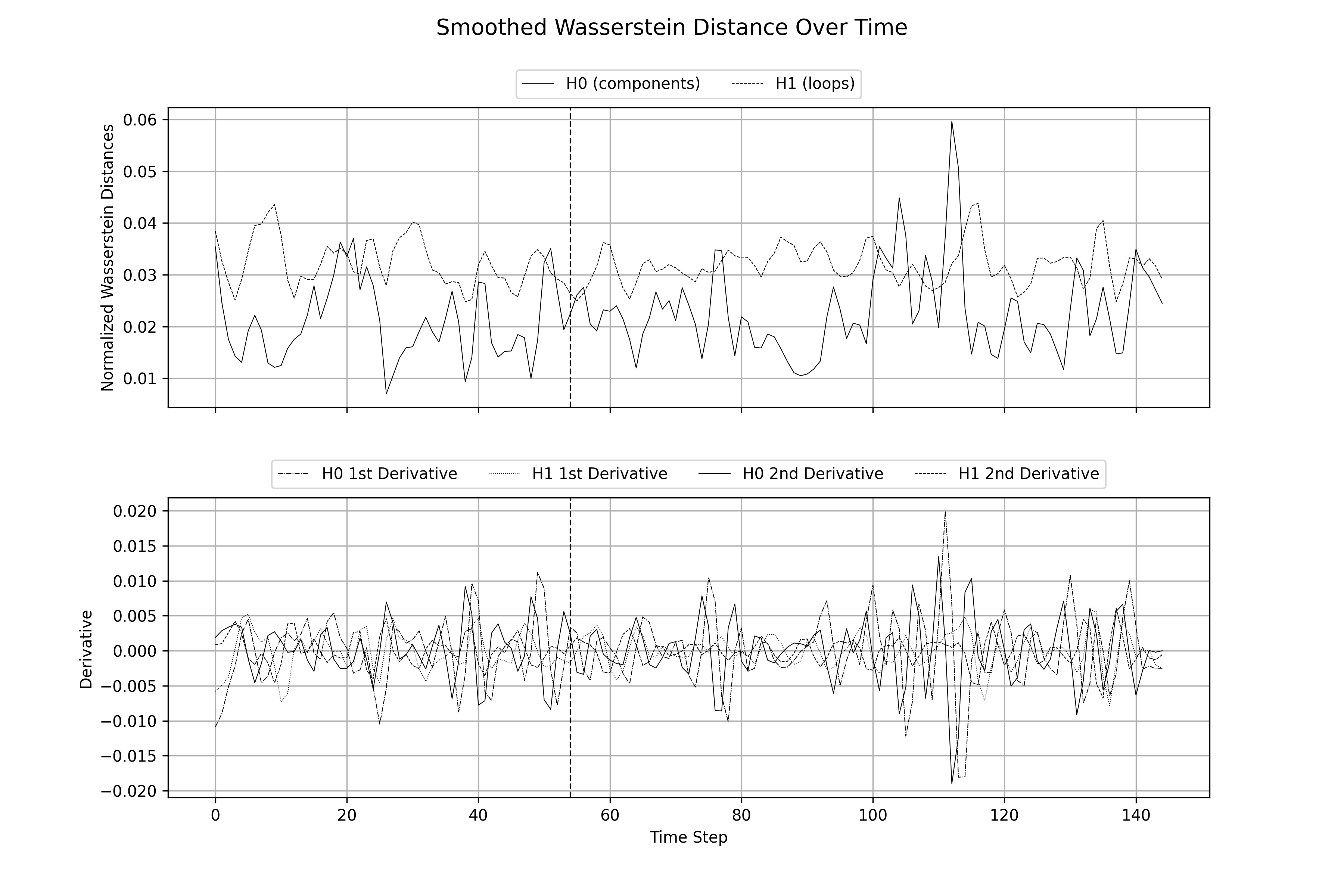}
    \caption{\textbf{February 24\textsuperscript{th} Dataset: Wasserstein Distance.} Smoothed, normalized Wasserstein distances over time (top panel) and their first and second temporal derivatives (bottom panel), computed from persistence diagrams in homology dimensions 0 ($H_0$, connected components) and 1 ($H_1$, loops). A clear spike in $H_0$ distance derivatives emerges just after the marked event (February 24\textsuperscript{th}, 2022), suggesting rapid topological change in the component structure of discourse. In contrast, $H_1$ distances remain comparatively stable, with only modest derivative fluctuations. This indicates that the semantic reorganization driven by the invasion was more pronounced at the level of narrative connectivity than in cyclical structure.}
    \label{fig:f_24_distance_derivatives}
\end{figure}

\begin{figure}[htbp]
    \centering
    \includegraphics[width=\linewidth]{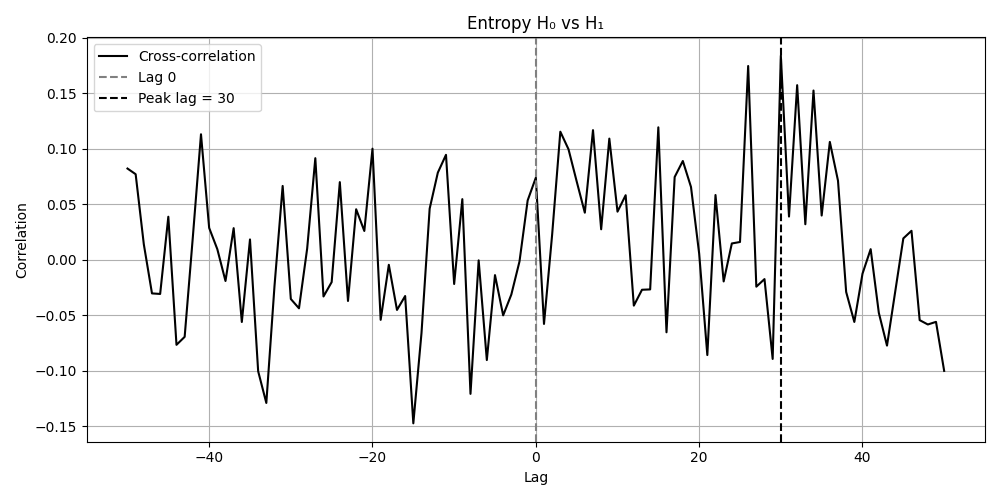}
    \caption{\textbf{February 24\textsuperscript{th} Dataset: Persistence Entropy Cross Correlation.} Cross-correlation between $H_0$ and $H_1$ persistence entropy time series. The peak correlation occurs at a lag of +30, indicating that changes in component-level entropy \textit{follow} changes in loop-level entropy by approximately 30 time steps. This suggests that disruptions in larger thematic or cyclical structures ($H_1$) may precede and influence the reorganization of more localized conceptual clusters ($H_0$), implying a top-down dynamic in narrative restructuring.}
    \label{fig:f_24_entropy_xcorr}
\end{figure}

\begin{figure}[htbp]
    \centering
    \includegraphics[width=\linewidth]{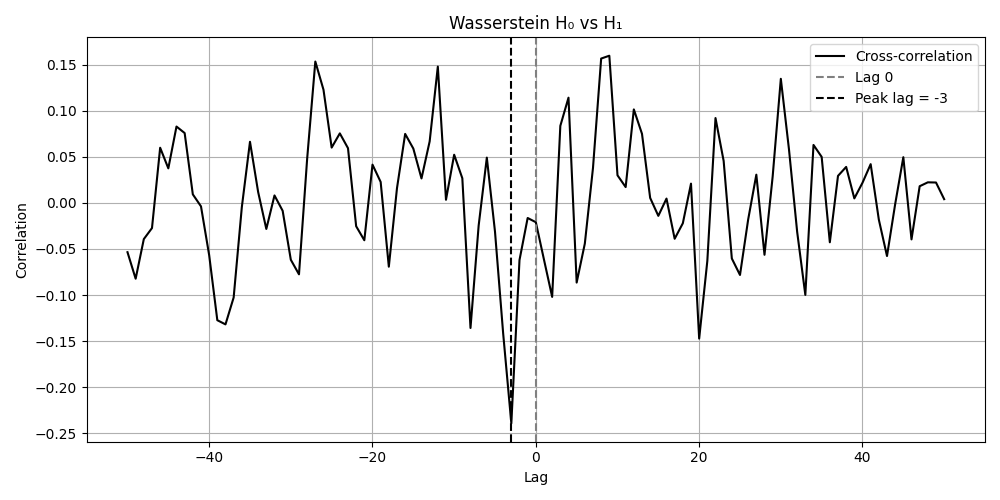}
    \caption{\textbf{February 24\textsuperscript{th} Dataset: Wasserstein Distance Cross Correlation.} Cross-correlation between $H_0$ and $H_1$ Wasserstein distances. The peak occurs at a lag of -3, indicating that shifts in connected component structure slightly precede those in loop structure. The small but negative lag supports a bottom-up interpretation of semantic reconfiguration, in which early disruptions to basic narrative connectivity eventually affect more complex relational motifs.}
    \label{fig:f_24_wasserstein_xcorr}
\end{figure}

\textbf{Topological Signal Analysis of the February 24\textsuperscript{th} Dataset.} In the February 24\textsuperscript{th} data set, the entropy analysis revealed a marked decline in $H_0$ entropy just prior to the event, followed by increased volatility and elevated derivative magnitudes—signaling a disruption in narrative cohesion and a transition to a more disordered semantic regime (Figure \ref{fig:f_24_entropy_derivatives}). In contrast, $H_1$ entropy remained relatively stable in magnitude but exhibited structural complexity that appeared to precede changes in $H_0$. Interestingly, cross-correlation analysis revealed that changes in $H_0$ entropy lagged those in $H_1$ by approximately $30$ time steps—reversing the typical pattern seen in other event-aligned datasets (Figure \ref{fig:f_24_entropy_xcorr}). This suggests that disruptions in higher-order thematic or framing structures (represented by $H_1$) preceded the breakdown and reorganization of local semantic clusters (represented by $H_0$). Such a top-down shift implies that abstract, global frames of interpretation—e.g., “invasion,” “sovereignty,” “war”—entered the discourse before fine-grained topics such as locations, actors, or tactical narratives were semantically assimilated. This temporal inversion may reflect the rapid imposition of interpretive narratives by global media and political actors, producing a cascading reconfiguration from macro-level themes down to micro-level content. The Wasserstein distance derivatives showed a sharp spike in $H_0$ immediately after the event, further confirming a reorganization in the discourse structure at the component level (Figure \ref{fig:f_24_distance_derivatives}). In contrast to the entropy lag, the Wasserstein cross-correlation showed a more modest lead of $-3$ for $H_1$ over $H_0$, indicating a partially consistent temporal ordering across metrics (Figire \ref{fig:f_24_wasserstein_xcorr}). These findings demonstrate the sensitivity of topological methods not only to structural disruptions but also to shifts in causal hierarchy—capturing how narrative power can propagate through different levels of semantic representation in response to major geopolitical events.

\subsection{May 25th, 2020}
\begin{figure}[htbp]
    \centering
    \includegraphics[width=\linewidth]{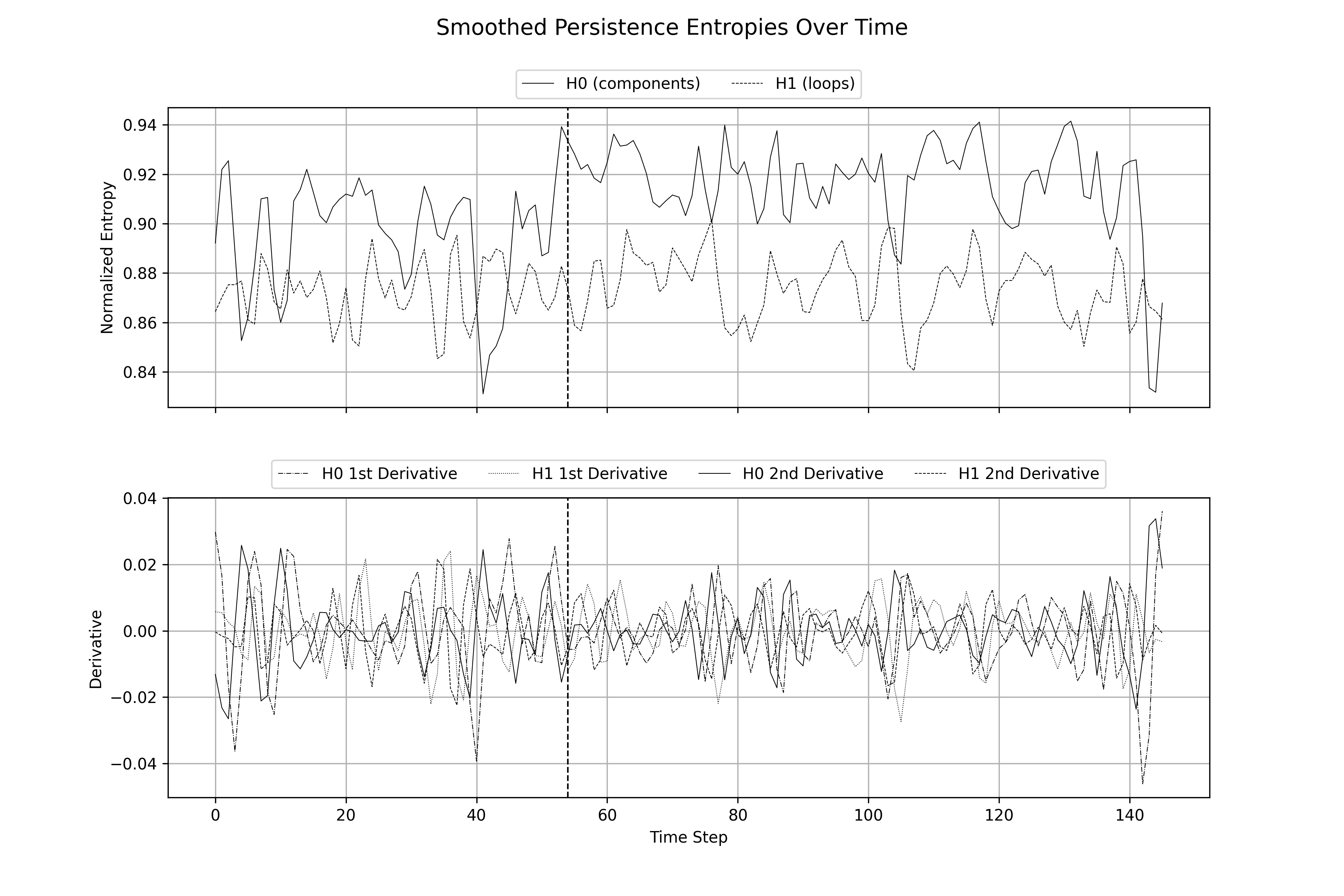}
    \caption{\textbf{May 25\textsuperscript{th} Dataset: Persistence Entropy.} Normalized persistence entropy traces (top panel) and their first and second derivatives (bottom panel) for $H_0$ and $H_1$. Following the event, both homology dimensions show increased volatility, but $H_0$ entropy rises more sharply and sustains elevated levels, suggesting an expansion in the number of short-lived narrative clusters. Derivative traces indicate abrupt topological acceleration in both dimensions, especially $H_0$, reflecting widespread narrative fragmentation and the emergence of new, unstable components in media discourse.}
    \label{fig:may_25_entropy_derivatives}
\end{figure}

\begin{figure}[htbp]
    \centering
    \includegraphics[width=\linewidth]{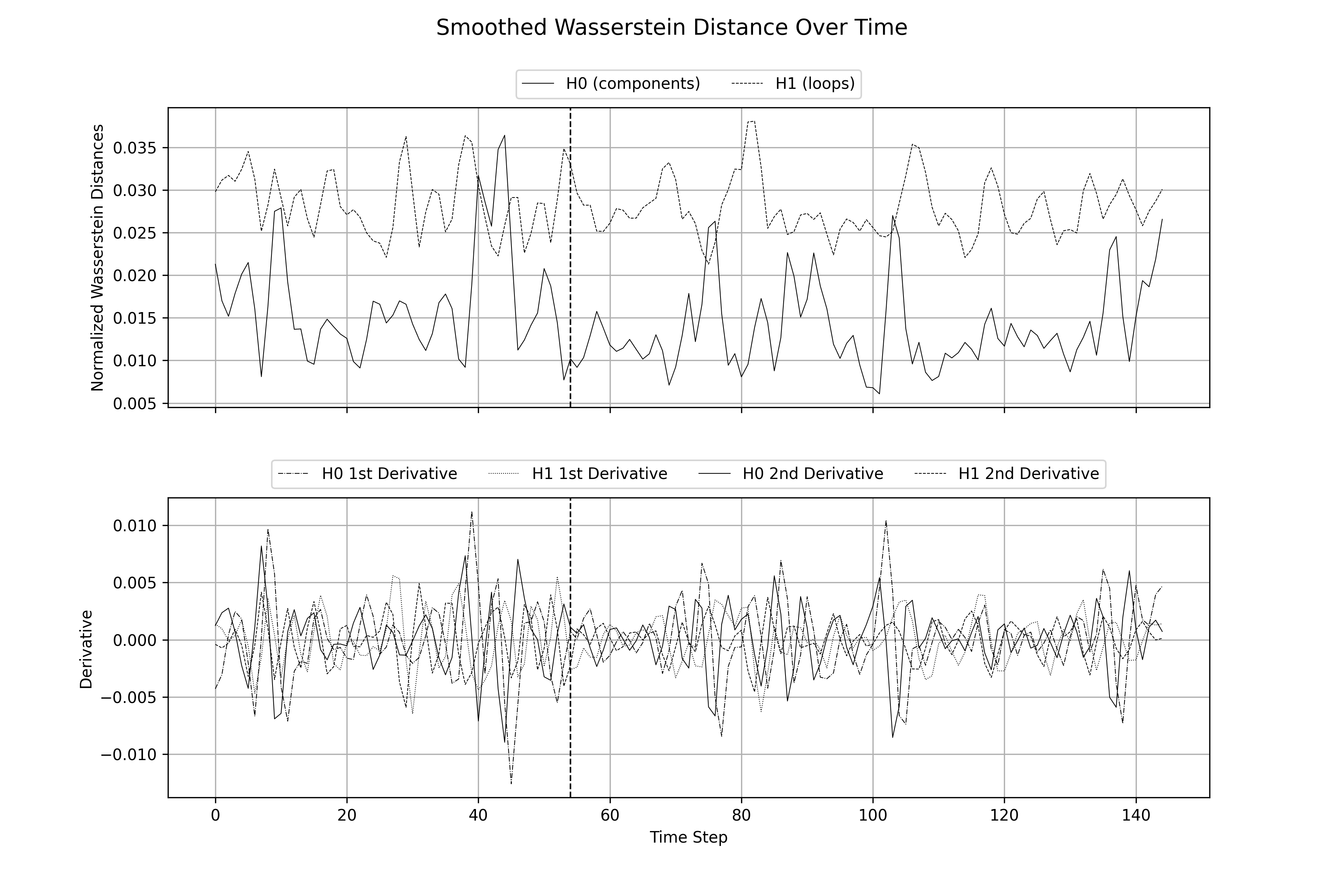}
    \caption{\textbf{May 25\textsuperscript{th} Dataset: Wasserstein Distance.} Smoothed, normalized Wasserstein distances (top panel) and their first and second derivatives (bottom panel) for persistence diagrams in $H_0$ (connected components) and $H_1$ (loops). A notable divergence emerges shortly after the event marker, where $H_0$ distances display a transient decrease in volatility, while $H_1$ shows relatively greater dynamism. This suggests a brief phase of semantic stabilization in component structure, followed by elevated cyclic variability—perhaps indicating a shift from structural consensus to a proliferation of contested narrative framings.}
    \label{fig:may_25_distance_derivatives}
\end{figure}

\begin{figure}[htbp]
    \centering
    \includegraphics[width=\linewidth]{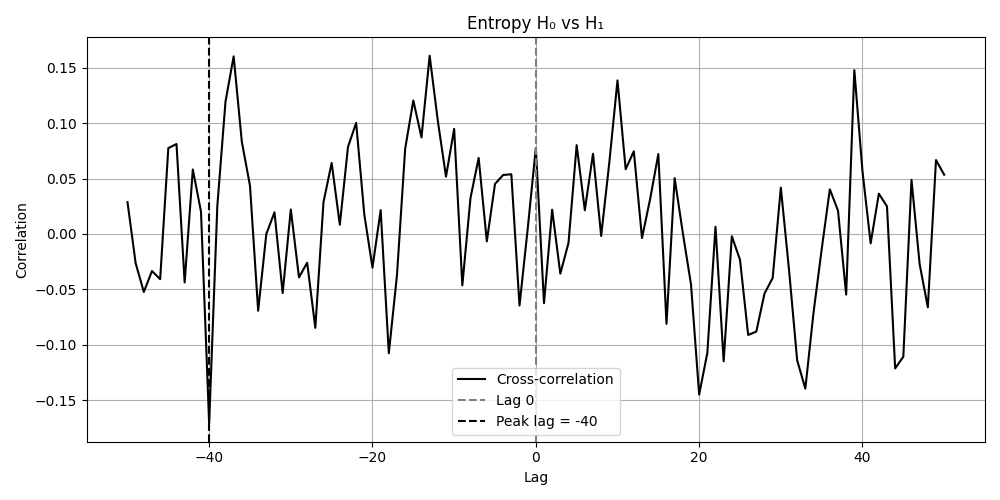}
    \caption{\textbf{May 25\textsuperscript{th} Dataset: Persistence Entropy Cross Correlation.} Cross-correlation between $H_0$ and $H_1$ persistence entropy time series. The peak occurs at lag $-40$, indicating that changes in component-level entropy precede those in loop-level entropy by approximately 40 time steps. This points to a sequential unfolding in narrative structure, where a breakdown in conceptual cohesion initiates longer-term disruptions in thematic framing and cyclic motifs.}
    \label{fig:may_25_entropy_xcorr}
\end{figure}

\begin{figure}[htbp]
    \centering
    \includegraphics[width=\linewidth]{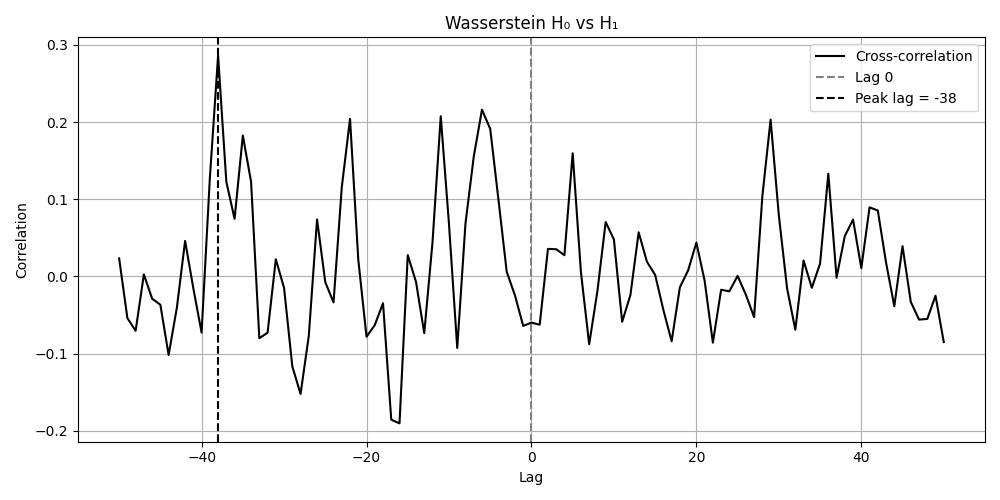}
    \caption{\textbf{May 25\textsuperscript{th} Dataset: Wasserstein Distance Cross Correlation.} Cross-correlation between $H_0$ and $H_1$ Wasserstein distances. The peak lag occurs at $-38$, supporting the same temporal hierarchy observed in the entropy traces. This suggests that reconfiguration in the global structure of semantic connectivity leads, rather than follows, changes in the cyclic organization of discourse. The substantial lag indicates a prolonged semantic cascade following the initial rupture.}
    \label{fig:may_25_wasserstein_xcorr}
\end{figure}

\textbf{Topological Signal Analysis of the May 25\textsuperscript{th} Dataset.} In the May 25\textsuperscript{th} dataset, $H_0$ entropy rose significantly after the event and exhibited sustained volatility, reflecting a disruption in semantic cohesion and the emergence of many short-lived components (Figure \ref{fig:may_25_entropy_derivatives}). In contrast, $H_1$ entropy and distances indicated a delayed response, suggesting slower reorganization of relational framing structures (Figure \ref{fig:may_25_distance_derivatives}). Cross-correlation analysis confirmed this temporal ordering: both entropy and Wasserstein distances peaked at large negative lags (-40 and -38, respectively), with $H_0$ leading $H_1$ (Figures \ref{fig:may_25_entropy_xcorr} and \ref{fig:may_25_wasserstein_xcorr}). These results suggest a bottom-up semantic disruption, where foundational narrative elements were destabilized first, precipitating more gradual shifts in complex thematic cycles. Persistent homology thus captures a clear, multi-timescale reconfiguration of discourse in the wake of a socially catalytic event.

\FloatBarrier

\section{Discussion}
Our analysis demonstrates that persistent homology provides a robust and interpretable framework for detecting structural transitions in public discourse. By tracing the evolution of topological features in daily semantic graphs, we identified sharp reorganizations in narrative structure that coincide with globally significant events, including the Russian invasion of Ukraine, the murder of George Floyd, the U.S. Capitol insurrection, and the Hamas-led invasion of Israel. Across all datasets, we observed characteristic patterns in both $H_0$ (connected components) and $H_1$ (loops), reflecting distinct phases of disruption, reorganization, and stabilization in the semantic topology of news narratives.

\subsection{Bottom-Up Reorganization}
A consistent finding across most events was the temporal precedence of changes in $H_0$ over $H_1$—both in persistence entropy and in Wasserstein distances. This lag structure suggests a bottom-up dynamic of semantic reconfiguration: ruptures in narrative cohesion (reflected in $H_0$) tend to precede shifts in higher-order relational motifs (captured by $H_1$). For example, in the aftermath of George Floyd’s murder, entropy cross-correlation peaked at a lag of $-40$, while Wasserstein distance peaked at $-38$, indicating that structural fragmentation in the basic units of discourse initiated a delayed cascade into more complex, thematic reorganization. A similar temporal pattern emerged in the Capitol insurrection and October 7\textsuperscript{th} datasets, reinforcing the notion that large-scale events propagate their semantic impact gradually—from the disintegration of shared referents to the restructuring of collective meaning.

\subsection{Event-Specific Dynamics and Inverted Causal Order}
Despite these shared dynamics, certain events exhibited striking deviations from the bottom-up pattern. Most notably, in the February 24\textsuperscript{th} dataset corresponding to the Russian invasion of Ukraine, cross-correlation analysis revealed that changes in $H_0$ entropy \textit{lagged} those in $H_1$ by approximately $+30$ time steps. This reversed ordering suggests a top-down semantic dynamic, wherein disruptions in abstract thematic or framing structures (captured by $H_1$) preceded the reconfiguration of local narrative clusters (captured by $H_0$). Such a pattern may reflect the rapid imposition of global interpretive frames—e.g., “invasion,” “war,” “sovereignty”—before granular, fact-level reporting and narrative alignment occurred. This inversion stands in contrast to the other datasets and underscores the potential of topological methods to uncover variation in the causal hierarchy of semantic change. 

Other events, such as May 25\textsuperscript{th}, showed more protracted entropy lags and sustained volatility, consistent with ongoing contestation and fragmented public narratives. These differences may reflect the nature of the events themselves (e.g., rapid rupture versus prolonged unrest), the pace of media coverage, or differences in audience segmentation and epistemic coherence.

\subsection{Stability and Volatility}
Derivative analysis of both entropy and Wasserstein distances revealed key turning points in the topological evolution of discourse. Spikes in first-order derivatives typically marked the onset of structural change, while inflection points in second-order derivatives signaled transitions into new regimes—often interpreted as stabilization or plateauing following the event. This pattern was especially clear in the February 24\textsuperscript{th} dataset, where post-invasion volatility gradually resolved into a new, persistent discourse structure. Conversely, the May 25\textsuperscript{th} dataset showed prolonged topological instability, consistent with a sustained period of social unrest and competing narrative framings.

\subsection{Implications for Event Detection and Discourse Analysis}
These findings suggest that topological signals offer a sensitive and temporally nuanced tool for event detection, capable of identifying not only the onset but also the unfolding and aftershocks of semantic reconfiguration. Importantly, our approach is entirely unsupervised: it requires no prior knowledge of the event, no keyword dictionaries, and no labeled training data. This enables the detection of emergent or unanticipated phenomena that may be missed by conventional methods focused solely on lexical salience or sentiment. Furthermore, the use of persistent homology enables a scale-invariant view of structure, capturing both localized disruptions and systemic transitions across time.

\section{Conclusions and Future Work}
This study presents a novel application of topological data analysis to the evolving structure of media discourse, offering a mathematically grounded method for detecting systemic shifts in public narratives. By embedding daily co-occurrence graphs of noun phrases and computing their persistent homology, we identified moments of sharp semantic reorganization that coincided with high-salience geopolitical and social events. The combined use of Wasserstein distance, persistence entropy, and cross-correlation analysis enabled us to detect both immediate disruptions and delayed narrative cascades across different homological dimensions.

Our results generally show that $H_0$ features—representing the emergence and dissolution of narrative components—respond first to external shocks, while $H_1$ features—capturing higher-order semantic loops—often shift more gradually. This temporal structure suggests that foundational narrative elements destabilize prior to the reorganization of complex thematic framings. However, the exception observed on February 24\textsuperscript{th}—in which $H_1$ entropy led changes in $H_0$—demonstrates that this ordering is not universal. In cases of rapid global framing, high-level interpretive schemas may precede factual narrative realignment, reversing the usual semantic cascade. This underscores the value of persistent homology not only as a detector of disruption but also as a probe into the semantic directionality of narrative change.

\subsection{Future Work}
Several avenues for future research remain. First, we aim to extend this framework to a broader class of events, including financial crises, natural disasters, and political transitions, to test its applicability across different domains of collective sense-making. Second, integrating multilingual corpora could reveal whether structural shifts propagate differently across cultural or linguistic boundaries. Third, coupling persistent homology with sentiment trajectories or topic models could enrich our understanding of how affective and thematic content evolve in tandem with discourse topology.

We also envision applications of this method to real-time event detection and monitoring. By continuously updating topological indicators on incoming media streams, it may be possible to develop early-warning systems that flag semantic discontinuities before they manifest in conventional reporting. In addition, further exploration of higher-dimensional homological features ($H_2$ and above) may provide insight into even more intricate patterns of discourse entanglement.

Ultimately, this work contributes to the growing toolkit of computational social science, demonstrating how topological methods can illuminate the shape of public thought as it is formed, fractured, and reassembled in response to the shocks of history.

\medskip

\bibliographystyle{unsrtnat}
\bibliography{references}

\end{document}